\documentclass[oneside,english]{elsartdense}
\usepackage[T1]{fontenc}
\usepackage[latin9]{inputenc}
\usepackage{amssymb}
\usepackage{esint}

\makeatletter

\usepackage{babel}
\makeatother

\begin{document}

\begin{frontmatter}

\title{Quantum of area from gravitation on complex octonions}

\author{Jens K\"oplinger}

\address{105 E Avondale Dr, Greensboro, NC 27403, USA}

\ead[url]{http://www.jenskoeplinger.com/P}

\begin{abstract}
Using spin 1/2 particle elastic scattering on a fixed target, in a
$1/\left|\vec{x}\right|$ potential on Euclidean metric, a minimum
scattering cross section appears from the spin contribution. Interpreted
as semi-classical limit of an earlier proposed operator formulation
of four dimensional Euclidean quantum gravity, using non-associative
complex octonion algebra, this is understood as an area quantum for
the observation: Existence of finite (quantized) charges yields a
minimum area element of interaction. This suggests that models built
on the fundamental assumption of a quantized area element (namely
as in Loop Quantum Gravity) may in principle approximate General Relativity
in the non-quantum limit.
\end{abstract}
\end{frontmatter}

\section{Overview}

This paper connects to recent advances towards quantum theory on octonionic
algebra \cite{DzhunushalievObsAndUnobs2007,DzhunushalievToyModels2007,DzhunushalievNonAssocHidden2007,DzhunushalievAHiddenNonassoc,GogberashviliOctonionicGeom2006,GogberashviliOctonionicDirac2006,GogberashviliOctonionicElectrodyn2006,Koepl2006DiracEqn,KoeplSigGravity2007,KoeplHypRelativity2007,KoeplGravEMConSed2007,DrayManogueQuatSpin1999,DrayManogueDimReduction1998,TolanOzdasTanisli2006,CandemirEtAlHypOctProcaMax2008,Gogberashvili2008RotationsOct},
and supplies the result for lowest-order quantum gravitational Coulomb
scattering of a spin-1/2 particle on a static point target, in a four
dimensional Euclidean quantum gravity model on complex octonions.
The resulting elastic scattering cross section has properties that
one might expect from gravity: Backscattering is enhanced, the cross
section does not vanish for increasing energies, and the general weakness
of quantum gravitational interaction poses challenges for experimental
validation.

A new effect also appears: The observed elastic scattering cross section
may not be smaller than a finite area element, which only depends
on the charges of particle and target. It is concluded that a quantized
charge implies a minimum area of interaction in this model, a quantum
of area.

In principle, this observation is consistent with approaches to gravity
on quantized length elements, namely Loop Quantum Gravity, and it
is pointed out that here it originates from a model that approximates
General Relativity on the non-quantum scale.

\section{Spin 1/2 Coulomb scattering in lowest order}

The gravitational elastic scattering cross section $d\sigma$ will
be calculated in lowest order, for a massive spin 1/2 particle on
a static target, as follows:

\begin{enumerate}
\item Within the generally non-associative complex octonion formulation
of electromagnetism and gravity, the {}``circular Dirac equation''
from \cite{KoeplSigGravity2007}, equation (26), will be identified
as the associative subalgebra, modeling four dimensional Euclidean
quantum gravity.
\item A static $1/\left|\vec{x}\right|$ potential will be supplied according
to \cite{KoeplGravEMConSed2007} equation (13), to yield the cross
section $d\sigma$ without spin.
\item The spin contribution will be calculated, and shown to increase backscattering.
A spin-flip will not occur.
\item In order to quantify this effect for the experiment, the semi-classical
projection from Euclidean into Minkowskian observer space-time geometry
will be executed, according to \cite{KoeplHypRelativity2007} proposition
4 ({}``\emph{NatAliE equations}'').
\item In the result, forward- and backscattering effects will be in qualitative
agreement with \emph{General Relativity} expectations. The interaction
yields a minimum area element.
\end{enumerate}

\subsection{The associative operator on complex octonions suitable for gravity}

Complex octonions are octonions with complex number coefficients,
$\mathbb{C}\otimes\mathbb{O}$, and are synonymous with {}``conic
sedenions'' in some of the referenced publications \cite{Carmody1988CircAndHyp,Carmody1997CircAndHypFurtherRes}.
In the current paper, terminology will be used that is more common
with contemporary uses, i.e.~complex octonions (instead of {}``conic
sedenions''), split-octonions (instead of {}``hyperbolic octonions''),
and octonions (instead of {}``circular octonions'').

Complex octonions will be expressed to a basis $b_{\mathbb{C}\otimes\mathbb{O}}:=\left\{ 1,i_{1},\ldots,i_{7},\quad i_{0},\epsilon_{1},\ldots,\epsilon_{7}\right\} $,
where $b_{\mathbb{O}}:=\left\{ 1,i_{1},\ldots,i_{7}\right\} $ is
a chosen octonion basis, and the remaining elements $\left\{ i_{0},\epsilon_{1},\ldots,\epsilon_{7}\right\} $
result from multiplication with the commutative and associative $i_{0}$,
i.e.~$\left\{ 1,i_{0}\right\} \otimes\left\{ 1,i_{1},\ldots,i_{7}\right\} \equiv\mathbb{C}\otimes\mathbb{O}$,
with $i_{0}^{2}=-1$ and $\epsilon_{n}^{2}=+1$. For a complete multiplication
table see e.g.~\cite{DzhunushalievToyModels2007} (table 1); for
octonion algebra in general see e.g.~\cite{BaezOctonions2002}.

For a spin-1/2 particle of mass $m$, charge $e$, mixing angle $\alpha$,
in a field field $A_{\mu}$ ($\mu=0\ldots3$), the terms from \cite{KoeplGravEMConSed2007},
relations (17) through (23),\begin{eqnarray}
\nabla_{\mathrm{Q1}} & := & \left(0,\partial_{0},0,0,0,0,0,0,\quad0,eA_{0},0,0,0,0,0,0\right),\\
\nabla_{\mathrm{Q2}} & := & \left(0,0,0,0,0,\partial_{3},-\partial_{2},\partial_{1},\quad0,0,0,0,0,eA_{3},-eA_{2},eA_{1}\right),\\
\nabla & := & \nabla_{\mathrm{Q1}}+\exp\left(i_{0}\alpha\right)\nabla_{\mathrm{Q2}},\\
\Psi_{\mathrm{Q1}} & := & \left(\psi_{0}^{\mathrm{r}},\psi_{0}^{\mathrm{i}},\psi_{1}^{\mathrm{r}},\psi_{1}^{\mathrm{i}},0,0,0,0,\quad0,0,0,0,0,0,0,0\right),\\
\Psi_{\mathrm{Q2}} & := & \left(0,0,0,0,-\psi_{2}^{\mathrm{r}},\psi_{2}^{\mathrm{i}},\psi_{3}^{\mathrm{r}},\psi_{3}^{\mathrm{i}},\quad0,0,0,0,0,0,0,0\right),\\
\Psi & := & \Psi_{\mathrm{Q1}}+\exp\left(i_{0}\alpha\right)\Psi_{\mathrm{Q2}},\end{eqnarray}
allow to formulate an equation of motion of a spin-1/2 particle in
a gravitational and electromagnetic field as:\begin{equation}
\left(\nabla-m\right)\Psi=0.\label{eq:DiracEqnGREMFieldCon16}\end{equation}
While this relation is in general non-associative, the cases $\alpha=\pm n\pi/2$
(with $n=0,1,2,\ldots$) can be reduced to associative matrix expressions:
$\alpha=\pi/2$ ($\pm n\pi$) becomes the Dirac equation with electromagnetic
field, and $\alpha=0$ ($\pm n\pi$) the proposed counterpart with
gravitational field. It is required to find such associative subalgebras
within a generally non-associative formulation, because the non-associative
parts of that formulation describe unobservable parameters that cannot
be measured in principle \cite{DzhunushalievNonAssocHidden2007,DzhunushalievAHiddenNonassoc}.

For all known particles, the mixing angle $\alpha$ is most nearly
$\alpha\approx\pi/2+m/e$, i.e.~it deviates from $\pi/2$ only in
the order of magnitude of the relative strength between gravitational
and electromagnetic force of that particle. For general $\alpha$,
however, all three parameters ($e$, $m$, and $\alpha$) are required
to characterize a particle: $e$ becomes a generalized charge that
is independent from its type of interaction (electromagnetic or gravitational),
$\alpha$ parametricizes the type of interaction, and $m$ becomes
the invariant property\begin{eqnarray}
m=\left|p\right| & = & \sqrt[4]{E^{4}+\left|\vec{p}\right|^{4}+2E^{2}\left|\vec{p}\right|^{2}\left(\cos^{2}\alpha-\sin^{2}\alpha\right)},\end{eqnarray}
with $E$ and $\vec{p}$ the Fourier coefficients of $t\equiv x_{0}$
and $\vec{x}$ respectively.

In order to calculate the effects from quantum gravity, the case $\alpha=0$
will now be examined, to yield the {}``circular Dirac equation''
from \cite{KoeplSigGravity2007}, relation (26), with field $A_{\nu}$
as\textbf{}%
\footnote{All summations are spelled out, all indices are lower indices, and
a metric tensor would be written explicitly if present. The chosen
notation avoids a potential ambiguity between Euclidean ($\delta_{\mu\nu}$)
and Minkowskian space-time ($\eta_{\mu\nu}$).%
}:\begin{equation}
\left[\sum_{\nu=0}^{3}\beta_{\nu}\left(i\partial_{\nu}-eA_{\nu}\right)-m\right]\Psi=0.\label{eq:circDiracEquationWField}\end{equation}
The attribute {}``circular'' refers to its underlying four-dimensional
Euclidean space-time, i.e.\begin{eqnarray}
\frac{1}{2}\left(\beta_{\mu}\beta_{\nu}+\beta_{\nu}\beta_{\mu}\right) & = & \delta_{\mu\nu},\label{eq:BetaMatricesMetricTensorIsUnity}\\
\sum_{\mu=0}^{3}\sum_{\nu=0}^{3}\beta_{\mu}\beta_{\nu}p_{\mu}p_{\nu} & = & \sum_{\nu=0}^{3}p_{\nu}^{2}=E^{2}+\left|\vec{p}\right|^{2}=m^{2}\label{eq:SumBetaBetaPPEqyalsP2}\end{eqnarray}

Such particle would only interact gravitationally, could travel in
excess of the speed of light, and is therefore hypothetical. However,
calculating its scattering cross section allows to subsequently perform
the semi-classical projection on a Minkowskian observer, to model
a realistic particle according to the \emph{NatAliE Equations} program
from \cite{KoeplHypRelativity2007} proposition 4, and therofore to
compare the prediction with \emph{General Relativity}.

\subsection{Introducing a static Coulomb potential}

A static Coulomb potential\begin{eqnarray}
A_{0}:=\frac{Ze}{\left|\vec{x}\right|} & \qquad & A_{i}=0\qquad\left(i=1,2,3\right)\end{eqnarray}
is introduced into (\ref{eq:circDiracEquationWField}), where $Ze$
denotes the generalized charge of the central particle, expressed
in multiples of the incoming particle's generalized charge $e$. The
$\beta$ matrix relations (\ref{eq:BetaMatricesMetricTensorIsUnity})
and (\ref{eq:SumBetaBetaPPEqyalsP2}) are closely related to the electromagnetic
case, and allow to obtain the transition matrix element $S_{\mathrm{fi}}$
in lowest order, for elastic scattering, accordingly as:\begin{eqnarray}
S_{\mathrm{fi}} & = & im\int d^{4}x\,\psi_{\mathrm{f}}^{*}\frac{Ze}{\left|\vec{x}\right|}\psi_{\mathrm{i}}\end{eqnarray}
Separating the spin effects into a factor $\mathcal{U}$ to be calculated
later, the remaining parts of the wave functions $\Psi_{1/2}^{\pm}$,
from \cite{KoeplSigGravity2007} relations (20) through (23), have
identical properties as in the electromagnetic case, and yield in
the textbook result:\begin{eqnarray}
S_{\mathrm{fi}} & = & -i\frac{mZe^{2}}{V\sqrt{E_{\mathrm{i}}E_{\mathrm{f}}}}\,\mathcal{U}\,\delta\left(E_{\mathrm{i}}-E_{\mathrm{f}}\right)\int d^{3}x\, e^{i\vec{q}\vec{x}}\frac{1}{\left|\vec{x}\right|}.\label{eq:SfiForOneOverXPotential}\end{eqnarray}
Here, $E_{\mathrm{i}}$ and $E_{f}$ denote the initial and final
energy of the particle being scattered, $V$ is the unit volume of
interaction, and $\vec{q}$ is the change in momentum, $\vec{p_{\mathrm{i}}}-\vec{p_{\mathrm{f}}}$.
The well-known result for such $1/\vec{x}$ potential is:\begin{eqnarray}
\frac{d\sigma}{d\Omega} & = & \frac{Z^{2}e^{4}m^{2}}{4\left|\vec{p}\right|^{4}\sin^{4}\left(\theta/2\right)}\,\mathcal{U}^{2}.\label{eq:XSectionWOSpin}\end{eqnarray}

\subsection{Spin contribution}

\label{sub:SpinContribution}Introducing the spin contribution to
the circular Dirac equation, on Euclidean space-time $m^{2}=E^{2}+\left|\vec{p}\right|^{2}$,
requires the normalized eigenstates of \cite{KoeplSigGravity2007}
relations (20) through (23). Coordinates are rotated such that the
incoming particle will be travelling in $x_{1}$ direction, and scattering
happens in the $\left(x_{1},x_{2}\right)$ plane. As the entire calculation
is independent of the orientation of the time axis, particles $\Psi_{1}^{\pm}$
yield the same result as anti-particles $\Psi_{2}^{\pm}$.

For particles, the normalized eigenstates become:\begin{eqnarray}
\left|+\right\rangle :=u_{\mathrm{f/i}}^{+}=\sqrt{\frac{m+E}{2m}}\left(\begin{array}{c}
1\\
0\\
0\\
\left(-p_{1}-ip_{2}\right)/\left(m+E\right)\end{array}\right) & \qquad\left|-\right\rangle := & u_{\mathrm{f/i}}^{-}=\sqrt{\frac{m+E}{2m}}\left(\begin{array}{c}
0\\
1\\
\left(-p_{1}+ip_{2}\right)/\left(m+E\right)\\
0\end{array}\right)\end{eqnarray}
Spin flip may not occur, as expected:\begin{eqnarray}
\left\langle +\right|\beta_{0}\left|-\right\rangle  & = & \left(u_{\mathrm{f}}^{+}\right)^{*}\beta_{0}u_{\mathrm{i}}^{-}=0=\left\langle -\right|\beta_{0}\left|+\right\rangle \label{eq:NoSpinFlip}\end{eqnarray}
For unchanged spin $\left|+\right\rangle $ we have:\begin{eqnarray}
\left\langle +\right|\beta_{0}\left|+\right\rangle  & = & \left(u_{\mathrm{f}}^{+}\right)^{*}\beta_{0}u_{\mathrm{i}}^{+}\\
 & = & \frac{m+E}{2m}\left(1-\frac{\left(p_{\mathrm{f},1}-ip_{\mathrm{f},2}\right)\left(p_{i,1}+ip_{\mathrm{i},2}\right)}{\left(m+E\right)^{2}}\right).\end{eqnarray}

In terms of total momentum $\left|\vec{p}\right|$ and scattering
angle $\theta$, the relations\begin{eqnarray}
p_{\mathrm{f},1}p_{\mathrm{i},1}+p_{\mathrm{f},2}p_{\mathrm{i},2} & = & \vec{p_{\mathrm{f}}}\vec{p_{\mathrm{i}}}=\left|\vec{p}\right|^{2}\cos\,\theta,\\
p_{\mathrm{f},1}p_{\mathrm{i},2}-p_{\mathrm{f},2}p_{\mathrm{i},1} & = & \left(\vec{p_{\mathrm{f}}}\times\vec{p_{\mathrm{i}}}\right)_{3}=-\left|\vec{p}\right|^{2}\sin\,\theta,\end{eqnarray}
yield:\begin{eqnarray}
\left(p_{\mathrm{f},1}-ip_{\mathrm{f},2}\right)\left(p_{i,1}+ip_{\mathrm{i},2}\right) & = & \left|\vec{p}\right|^{2}\left(\cos\,\theta-i\,\sin\,\theta\right).\end{eqnarray}
Because $\left|\left\langle +\right|\beta_{0}\left|+\right\rangle \right|=\left|\left\langle -\right|\beta_{0}\left|-\right\rangle \right|$
due to symmetry, the total spin contribution is:\begin{eqnarray}
\mathcal{U}^{2} & = & \left|\left\langle +\right|\beta_{0}\left|+\right\rangle \right|^{2}\\
 & = & \left|\frac{\left(m+E\right)^{2}-\left|\vec{p}\right|^{2}\left(\cos\,\theta-i\,\sin\,\theta\right)}{2m\left(m+E\right)}\right|^{2}\\
 & = & \frac{\left(m+E\right)^{4}+\left|\vec{p}\right|^{4}-2\left(m+E\right)^{2}\left|\vec{p}\right|^{2}\cos\,\theta}{4m^{2}\left(m+E\right)^{2}}.\end{eqnarray}
Using $\cos\,\theta=1-2\sin^{2}\left(\theta/2\right)$ and $\left|\vec{p}\right|^{4}=\left(m^{2}-E^{2}\right)^{2}=\left(m+E\right)^{2}\left(m-E\right)^{2}$,
this becomes:\begin{eqnarray}
\mathcal{U}^{2} & = & \frac{1}{4m^{2}}\left(\left(m+E\right)^{2}+\left(m-E\right)^{2}-2\left|\vec{p}\right|^{2}\right)+\frac{\left|\vec{p}\right|^{2}}{m^{2}}\sin^{2}\frac{\theta}{2}\\
 & = & \frac{1}{4m^{2}}\left(2m^{2}+2E^{2}-2\left|\vec{p}\right|^{2}\right)+\frac{\left|\vec{p}\right|^{2}}{m^{2}}\sin^{2}\frac{\theta}{2}\\
 & = & \frac{E^{2}}{m^{2}}+\frac{\left|\vec{p}\right|^{2}}{m^{2}}\sin^{2}\frac{\theta}{2}=\frac{E^{2}}{m^{2}}\left(1+\left|\vec{v}\right|^{2}\sin^{2}\frac{\theta}{2}\right).\end{eqnarray}

The spin contribution yields a significant deviation from the electromagnetic
case, for this (hypothetical) interaction on purely Euclidean space-time:
Backscattering $\theta=\pi$ is enhanced (rather than suppressed),
and increasing kinetic energy of the incoming particle further enhances
the elastic cross section (instead of reducing it).

It is noted, however, that no such particle is known to exist, as
it would be interacting solely through gravitation (and no other force),
and would not be bound to the speed limit of light, or Heisenberg
uncertainty. The following section will approximate how known particles
would be affected by such a force.

\subsection{Semi-classical limit for comparison with \emph{General Relativity}}

\label{sub:SemiClassicalLimit}Solving the complex octonion equation
of motion (\ref{eq:DiracEqnGREMFieldCon16}) in the general case,
for a spin 1/2 particle in electromagnetic and gravitational fields,
requires mathematical methods yet to be determined. Argumentation
has been proposed in \cite{DzhunushalievObsAndUnobs2007,DzhunushalievToyModels2007,DzhunushalievNonAssocHidden2007},
that traditional quantum mechanical observables result from the associative
subalgebras of an otherwise non-associative, wider algebraic space.
It therefore appears feasible to use above calculations, as obtained
from an associative formulation on Euclidean space-time, and project
it onto the human observer space-time, which is Minkowskian.

This follows the same semi-classical limit as executed in \cite{KoeplHypRelativity2007},
termed the {}``\emph{NatAliE }equations'' program, which was used
there to show that such approach yields \emph{General Relativity }for
large bodies. There, it was shown in relation (58), that the projected
$1/\left|\vec{x}\right|$ potential of a point mass $m$, at rest
at the coordinate origin, would be seen by a Minkowskian observer
with speed $\left|\vec{v}\right|$ relative to $m$:\begin{eqnarray}
h_{00}' & = & -2\frac{m}{\left|\vec{x}\right|}\frac{1+\left|\vec{v}\right|^{2}}{\sqrt{1-\left|\vec{v}\right|^{2}}}\end{eqnarray}
Namely, a factor $\left(1+\left|\vec{v}\right|^{2}\right)/\sqrt{1-\left|\vec{v}\right|^{2}}$
needed to be applied to the static $1/\left|\vec{x}\right|$ potential,
which led to the linearized Einstein field equations for gravity.

Applied to the current calculations for spin 1/2 particle scattering,
such factor yields a transition matrix element that cannot be expressed
anymore in closed form, as $\left|\vec{v}\right|^{2}$ is a non-trivial
function of space and time. The exact expression of $S_{\mathrm{fi}}$
from relation (\ref{eq:SfiForOneOverXPotential}) above is:\begin{eqnarray}
\left.S_{\mathrm{fi}}\right|_{\mathrm{hyp}} & = & -i\frac{mZe^{2}}{V\sqrt{E_{\mathrm{i}}E_{\mathrm{f}}}}\,\mathcal{U}\,\delta\left(E_{\mathrm{i}}-E_{\mathrm{f}}\right)\int d^{3}x\, e^{i\vec{q}\vec{x}}\left(\frac{1}{\left|\vec{x}\right|}\frac{1+\left|\vec{v}\right|^{2}}{\sqrt{1-\left|\vec{v}\right|^{2}}}\right).\label{eq:SfiModiedByProjection}\end{eqnarray}

It is possible, however, to approximate this expression for non-relativistic
$\left|\vec{v}\right|$ and determine the resulting cross section
in closed form, as will now be shown.

Using\begin{eqnarray}
\frac{1+\left|\vec{v}\right|^{2}}{\sqrt{1-\left|\vec{v}\right|^{2}}} & \approx & 1+\frac{3}{2}\left|\vec{v}\right|^{2}+\mathcal{O}\left(\left|\vec{v}\right|^{4}\right)\end{eqnarray}
and the non-relativistic\begin{eqnarray}
\left|\vec{v}\right|^{2} & \approx & \frac{2}{m}\left(E_{\mathrm{kin,i}}+\frac{Ze^{2}}{\left|\vec{x}\right|}\right),\end{eqnarray}
(where $E_{\mathrm{kin,i}}$ is the initial kinetic energy at $\left|\vec{x}\right|\rightarrow\infty$),
one can approximate:\begin{eqnarray}
\frac{1}{\left|\vec{x}\right|}\frac{1+\left|\vec{v}\right|^{2}}{\sqrt{1-\left|\vec{v}\right|^{2}}} & \approx & \frac{1}{\left|\vec{x}\right|}\left(1+\frac{3E_{\mathrm{kin,i}}}{m}\right)+\frac{1}{\left|\vec{x}\right|^{2}}\left(\frac{3Ze^{2}}{m}\right)+\mathcal{O}\left(\left|\vec{v}\right|^{4}\right).\end{eqnarray}

The $1/\left|\vec{x}\right|$ Coulomb potential is enhanced by a small
factor, and a new $1/\left|\vec{x}\right|^{2}$ term is present. With
$1/\left|\vec{q}\right|^{2}$ being the Fourier pair to $1/\left|\vec{x}\right|$,
the expression\begin{eqnarray}
\left.S_{\mathrm{fi}}\right|_{\mathrm{hyp}} & \approx & -i\frac{mZe^{2}}{V\sqrt{E_{\mathrm{i}}E_{\mathrm{f}}}}\,\mathcal{U}\,\delta\left(E_{\mathrm{i}}-E_{\mathrm{f}}\right)\int d^{3}x\, e^{i\vec{q}\vec{x}}\left(\frac{\mathcal{A}}{\left|\vec{x}\right|}+\frac{\mathcal{B}}{\left|\vec{x}\right|^{2}}\right),\label{eq:SfiProjectedApproximated}\\
\mathcal{A} & := & 1+\frac{3E_{\mathrm{kin}}}{m},\\
\mathcal{B} & := & \frac{3Ze^{2}}{m},\end{eqnarray}
can be solved in closed form, to yield a cross section:\begin{eqnarray}
\left.\frac{d\sigma}{d\Omega}\right|_{\mathrm{hyp}} & = & 4Z^{2}e^{4}m^{2}\left(\frac{\mathcal{A}^{2}}{\left|\vec{q}\right|^{4}}+\frac{\mathcal{B}^{2}}{\left|\vec{q}\right|^{2}}\right)\mathcal{U}^{2}\label{eq:XSectionWoSpin2}\\
 & = & Z^{2}e^{4}m^{2}\left(\frac{\mathcal{A}^{2}}{4\left|\vec{p}\right|^{4}\sin^{4}\left(\theta/2\right)}+\frac{\mathcal{B}^{2}}{\left|\vec{p}\right|^{2}\sin^{2}\left(\theta/2\right)}\right)\left(\frac{E^{2}}{m^{2}}+\frac{\left|\vec{p}\right|^{2}}{m^{2}}\sin^{2}\frac{\theta}{2}\right)\\
 & = & \frac{Z^{2}e^{4}}{m^{2}}\left[\mathcal{A}^{2}\frac{m^{2}E^{2}}{4\left|\vec{p}\right|^{4}\sin^{4}\left(\theta/2\right)}+\mathcal{C}^{2}\frac{m^{2}}{\left|\vec{p}\right|^{2}\sin^{2}\left(\theta/2\right)}+\mathcal{D}^{2}\right],\label{eq:XSectionWSpin}\end{eqnarray}
with $\mathcal{A},\mathcal{C},\mathcal{D}$ dimensionless and:\begin{eqnarray}
\mathcal{C}^{2} & := & \frac{\mathcal{A}^{2}}{4}+\mathcal{B}^{2}E^{2},\qquad\mathcal{D}^{2}:=\mathcal{B}^{2}m^{2}=9Z^{2}e^{4}.\end{eqnarray}

The effective strenght of this effect can be estimated by replacing
the generalized charges $e$ and $Ze$ with the respective particles'
masses.

\subsection{Discussion of the result}

Spin contribution, as well as projection from Euclidean geometry (governing
gravity) into Minkowskian observer geometry, in non-relativistic approximation,
provide additional terms $\mathcal{A}^{2}$ and $\mathcal{B}^{2}/\left|\vec{q}\right|^{2}$
which enhance backscattering. They also prevent the total cross section
from vanishing at higher energies. These results are in qualitative
agreement, e.g., with textbook results from scattering of a non-quantum
particle on a Black Hole: Increased backscattering is commonly refered
to as {}``glory'', and a Black Hole will always retain a finite
area of capture of the incoming particle, no matter how large $\left|\vec{p}\right|$
or $E$ are.

A new effect is apparent when taking the spin contribution into account
as well: Relation (\ref{eq:XSectionWSpin}) contains a constant factor
$\mathcal{D}=3Ze^{2}$, that is independent from the dynamic parameters
$\left|\vec{p}\right|$, $E$, and angular distribution $\theta$.
In other words, the observed minimum (quantized) generalized charge
$e$ of a particle implies an observed minimum (quantized) area element,
as any incoming planar wave $\exp\left[i\left(\vec{p}\vec{x}\pm Et\right)\right]$
is uniformly scattered, independently of wave parameters $\left|\vec{p}\right|$
and $E$.

The order of magnitude of the quantum of area, however, is discouragingly
small. Introducing SI units, and replacing the generalized charge
$e$ with particle masses $m$ (incoming) and $M$ (target) respectively,\begin{eqnarray}
\frac{Z^{2}e^{4}}{m^{2}} & \rightarrow & \frac{G^{2}M^{2}}{c^{4}},\qquad9Z^{2}e^{4}\rightarrow\frac{9G^{2}M^{2}m^{2}}{\hbar^{2}c^{2}},\\
G & \approx & 6.674\times10^{-11}\,\frac{\mathrm{m^{3}}}{\mathrm{kg\, s^{2}}},\qquad\hbar\approx1.055\times10^{-34}\,\frac{\mathrm{kg\, m^{2}}}{\mathrm{s}},\qquad c\approx2.998\times10^{8}\,\frac{\mathrm{m}}{\mathrm{s}},\end{eqnarray}
the expression for the quantum of area becomes:\begin{eqnarray}
\left.\frac{d\sigma}{d\Omega}\right|_{\mathrm{area}} & = & \frac{9G^{4}}{\hbar^{2}c^{6}}M^{4}m^{2}\approx\left[2.21\times10^{-23}\,\frac{\mathrm{m}^{2}}{\mathrm{kg}^{6}}\right]\, M^{4}m^{2}.\end{eqnarray}
In the example electron $m_{e}\approx9.109\times10^{-31}\,\mathrm{kg}$
on proton $m_{p}\equiv M\approx1.673\times10^{-27}\,\mathrm{kg}$,
this would approximate $\left.d\sigma/d\Omega\right|_{\mathrm{area}}\approx1.44\times10^{-190}\,\mathrm{m^{2}}$,
roughly corresponding to the area of a circle with radius $4\times10^{-96}\,\mathrm{m}$.
This radius compares to the Planck length roughly the same as the
Planck length compares to the visible universe, and therefore safely
removes the effect from being accessible directly, through any experiment.

\section{Summary and outlook}

In this paper, operator quantum gravity on four dimensional Euclidean
space-time has been evaluated, for elastic spin 1/2 particle scattering
on a static target, in lowest order. It was found that backscattering
is enhanced, and that the cross section does not disappear for high
energies. a minimum scattering cross section appears, which only depends
on the interacting particles' masses and charges. With all known matter
currently assumed to be built from spin 1/2 particles, and elastic
scattering in lowest order arguably the most fundamental type of interaction,
this suggests that it is possible in principle to build models of
quantum gravity on the assumption of a quantized area element, to
yield General Relativity in the large body, non-quantum limit.

Further investigation will determine whether an operator description
of forces other than gravity and electromagnetism is possible \cite{DzhunushalievAHiddenNonassoc},
whether octonionic spinors allow to model the fermion generations
of the Standard Model \cite{DrayManogueDimReduction1998}, and how
this may relate to the symmetry groups used in classical Quantum Field
Theory \cite{Gogberashvili2008RotationsOct}.

\begin{ack}
I am thankful towards Merab Gogberashvili and Vladimir Dzhunushaliev
for their support.
\end{ack}

\end{document}